\documentclass[structabstract]{aa}  
\usepackage{natbib}
\bibpunct{(}{)}{;}{a}{}{,} % to follow the A&A style

\usepackage{graphicx}
\usepackage{color}
\usepackage{rotating}
\usepackage{journals}
\usepackage{txfonts}
\begin{document}

   \title{Fossil groups in the Millennium simulation}

   \subtitle{Their environment and its evolution}

  \author{Eugenia D\'{\i}az-Gim\'enez\inst{1},
	   Ariel Zandivarez\inst{1},
           Robert Proctor\inst{2},
           Claudia Mendes de Oliveira\inst{2}
	   \and
	   L. Raul Abramo\inst{3}
          }

   \institute{IATE (CONICET-UNC) \& OAC (UNC). Laprida 854, C\'ordoba 5000. Argentina\\
              \email{eugeniadiazz@gmail.com, arielz@mail.oac.uncor.edu}
         \and
             IAG, USP. Rua do Mat\~ao 1226, S\~ao Paulo. Brazil\\
             \email{rproctor@astro.iag.usp.br, oliveira@astro.iag.usp.br}
          \and
             Instituto de F\'{\i}sica, Universidade de S\~ao Paulo. 
             CP 66318, CEP 051314-970, S\~ao Paulo. Brazil.\\
             \email{abramo@fma.if.usp.br} 
         }

   \date{Received Month XX, year; accepted XX XX, year }

% \abstract{}{}{}{}{} 
% 5 {} token are mandatory
 
  \abstract
  {Fossil systems are defined to be X-ray bright galaxy groups (or clusters) with a 
  two-magnitude difference between their two brightest galaxies within
  half the projected virial radius, and represent an
  interesting extreme of the population of galaxy agglomerations.
  However, the physical conditions and processes leading to their
  formation are still poorly constrained.  }
   {We compare the outskirts of fossil systems
with that of normal groups to understand whether environmental
conditions play a significant role in their formation. We
study the groups of galaxies in both, numerical simulations and 
observations.
   }
   {We use a variety of statistical tools including the spatial
cross-correlation function and the local density parameter
$\Delta_5$ to probe differences in the density and structure of
the environments of 'normal' and 'fossil' systems in the Millennium
simulation.  
   }
  {We find that the number density of galaxies surrounding fossil
systems evolves from greater than that observed around normal
systems at z=0.69, to lower than the normal systems by z=0.  Both
fossil and normal systems exhibit an increment in their otherwise radially
declining local density measure ($\Delta_5$) at distances of order
2.5 $r_{vir}$ from the system centre.  We show that this increment is
more noticeable for fossil systems than normal systems and demonstrate
that this difference is linked to the earlier formation epoch of fossil groups.
Despite the importance of the assembly time, we show that the
environment is different for fossil and non-fossil systems with similar masses and
formation times along their evolution. We also confirm that the physical characteristics
identified in the Millennium simulation can also be detected in SDSS
observations. }
{Our results confirm the commonly held belief that fossil systems
assembled earlier than normal systems but also show that the surroundings
of fossil groups could be responsible for the formation of their large magnitude gap. 
   }
%% {}leave it empty if necessary  

   \keywords{ Methods:statistical--
              Galaxies:clusters:general--
              Galaxies:evolution
               }

   \authorrunning{D\'{\i}az-Gim\'enez et al.}
   \maketitle

%
%________________________________________________________________
%%%%%%%%%%%%%%%%%%%%%%%%%%%%%%%%%%%%%%%%%%%%%%%%%%%%%%%%%%%%%%%%%%%%%%%%%%%%

\section{Introduction}
The hierarchical structure formation paradigm is successful in
predicting many of the properties of galaxy groups and
clusters. However, to-date, the specific conditions and processes that
lead to the formation of a 'fossil' system have still to be
unequivocally identified.

'Fossil' systems are defined as spatially extended X-ray sources with
an X-ray luminosity $L_X>10^{42} \ h_{50}^{-2} \ erg \ s^{-1}$ whose
optical counterpart is a bound system of galaxies with $\Delta
M_{12}>2$ mag, where $\Delta M_{12}$ is the difference in absolute magnitude in
R-band between the brightest and the second brightest galaxies located
within half the projected virial radius of the systems.  ($r_{vir}$)
\citep{Jones03}. Broadly speaking, this means that fossil systems
consist of a relatively isolated, luminous, early-type galaxy embedded
in a swarm of much smaller galaxies and an extended X-ray halo. These
systems may therefore be of considerable importance as the place of
formation of giant elliptical galaxies.

The question then naturally arises: how do these systems form?  One
suggested scenario is that fossil systems form when a group or cluster
remains undisturbed for a significant fraction of a Hubble
time. Within the context of hierarchical structure formation, this
means that fossil systems would have assembled their dark matter halos
earlier than normal groups, thus leaving enough time for $L^{\star}$
objects to merge into the central galaxy by dynamical friction. In other words,
any large, centrally located satellite galaxies would have been tidally
stripped, disrupted, and finally cannibalised by the brightest
galaxy, naturally producing the characteristic magnitude gap ($\ge
2 mag$). In the present epoch, the central regions of fossil systems
would therefore exhibit the observed lack of $L^{\star}$ galaxies, but
the fainter end of the luminosity function would remain intact,
because of the longer timescale of dynamical friction for low mass galaxies.

The next question to arise is then: which systems are most likely to
experience this kind of evolution? Much effort has been devoted to
answering this question. Beside minor differences in the definition of
fossil groups, the fossil group phenomenon has been studied from the
broad range of observational, analytical, numerical, and semi-analytical points of
view \citep{Vik99,Jones03,Donghia,Mendes06,cyp06,KPJ06,M06,vandenbosch07,
sales07,vonbenda08,Mendes08}. Most of these works have
been motivated by the questions of whether the large magnitude
difference in fossil groups implies that they are a distinct class of
objects or they simply represent a tail of the cluster
distribution. Since the number of fossil systems identified
observationally is rather low, N-body cosmological simulations are of
considerable importance, as they allow us to perform statistical analysis. The
largest cosmological numerical simulation presently available is the
"Millennium simulation" (\citealp{Springel+05}, hereafter MS).
When combined with semi-analytical models of galaxy formation, this
simulation constitutes a useful tool in addressing open
issues surrounding the formation of fossil systems. Three
important studies used this tool to analyse the evolution of
fossil groups. On the one hand, \cite{Dariush07} concluded that fossil
systems identified in the MS assembled a larger fraction of their
masses at higher redshifts than non-fossil groups.  Therefore, they
suggest that the most likely scenario for fossil groups is that they are
not a distinct class of object but simply examples of
groups/clusters that collapsed early. \cite{dariush10} then
suggested refinements to the fossil definition to enhance its
efficiency in detecting old systems.  On the other hand, \cite{diaz08}
studied the first ranked galaxies in the MS, comparing the merger
history of the central galaxies in fossil and normal systems. By
analysing central galaxies with equal stellar mass distributions, they
found that, despite the earlier assembly time of fossil systems, first
ranked galaxies in fossil groups assembled half of their final mass and
experienced their last major merger \emph{later} than their non-fossil
counterparts, which implies that they followed a different evolutionary pathway. 
Consensus has clearly yet to be reached regarding the nature of fossil
systems.

Few works have studied the environment of fossil systems or the
influence that it could have had on their formation, mainly because
there are too few known fossil groups to allow a reliable statistical
analysis. However, observationally, it has been suggested that fossil
systems inhabit under-dense regions \citep{Jones03}. On the other hand,
\cite{vonbenda08} used numerical simulations to conclude that there is no
tendency for fossil systems to be preferentially located in low
density environments. They also showed that many galaxy groups and
clusters may undergo a fossil phase in their lives, but may not
necessarily still be fossil systems at $z=0$ because of the infall of $L^{\star}$
galaxies from the large-scale environment.  However, we
note that their study was limited to groups of relatively low virial
masses ($1-5 \ \times \ 10^{13} \ h^{-1} \ {\cal M}_{\odot}$), which
might not encompass the characteristic X-ray emission of observationally
selected fossil systems \citep{Dariush07}. In addition, several works
on galaxy groups have shown that their virial masses and the assembly times 
strongly depend on their environment \citep{Gao05,Berlind06,Wetzel07,Jing07}.

In this work, we attempt to obtain a clearer picture of fossil systems
and their environments by analysing the surroundings of systems
extracted from the MS. Throughout, we refer to groups and
clusters that meet the 'fossil' criteria above as 'fossil systems' or
simply 'fossils' and those that do not meet the criteria as
'non-fossils' or 'normal' systems.

The layout of this paper is as follows: In Sect.~\ref{samples}, we
describe the construction of the sample of fossil and non-fossil
groups.  In Sect.~\ref{xi_cg}, we study the outskirts of these groups using
the two-point cross-correlation function, while in Sect.~\ref{delta},
we describe their environment by means of the local density profile.
The analysis of the differences between the environments of fossils and non-fossils 
are carried out in Sect.~\ref{differences}.  A comparison
with observationally selected fossil groups is performed in
Sect.~\ref{observations}, while we summarise our conclusions in
Sect.~\ref{conclusions}.

\begin{figure}
%  -->$\Delta M_{12}<0.5 mag$
\centering
{\includegraphics[width=9.0cm]{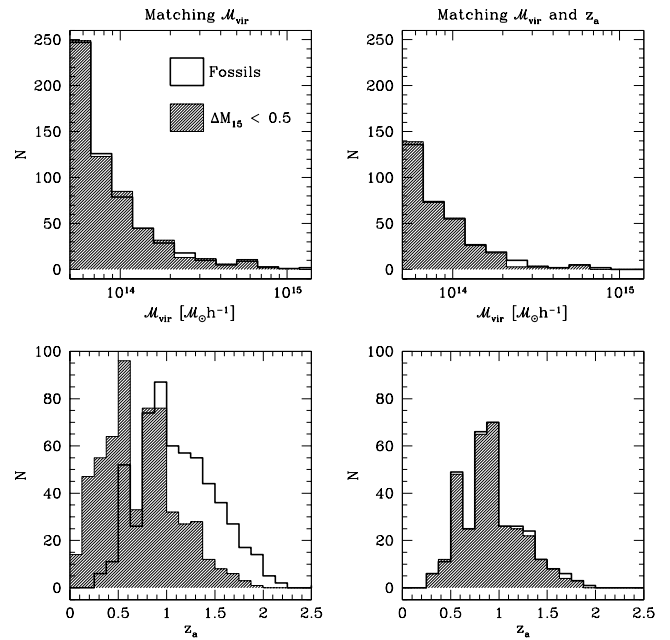}}
\caption{\emph{Upper panels} show the virial mass distributions of the
  samples of fossil (\emph{empty}) and $\Delta M_{12}<0.5 mag$
  (\emph{filled}) groups in the MS, while \emph{lower panels} show the
  assembly time distributions.  \emph{Left panels}: ``Mass-matched''
  samples.  Non-fossils that were selected to reproduce the same
  virial mass distribution of fossil groups in order to avoid a mass
  bias. \emph{Right panels}: ``Assembly-matched'' samples. Fossils and
  non-fossils were selected to have both similar virial mass
  \emph{and} assembly time distributions.  }
\label{masmatch}
\end{figure}

\begin{figure*}
% xi.sm --> xi2nT
\centering
{\includegraphics[width=15cm]{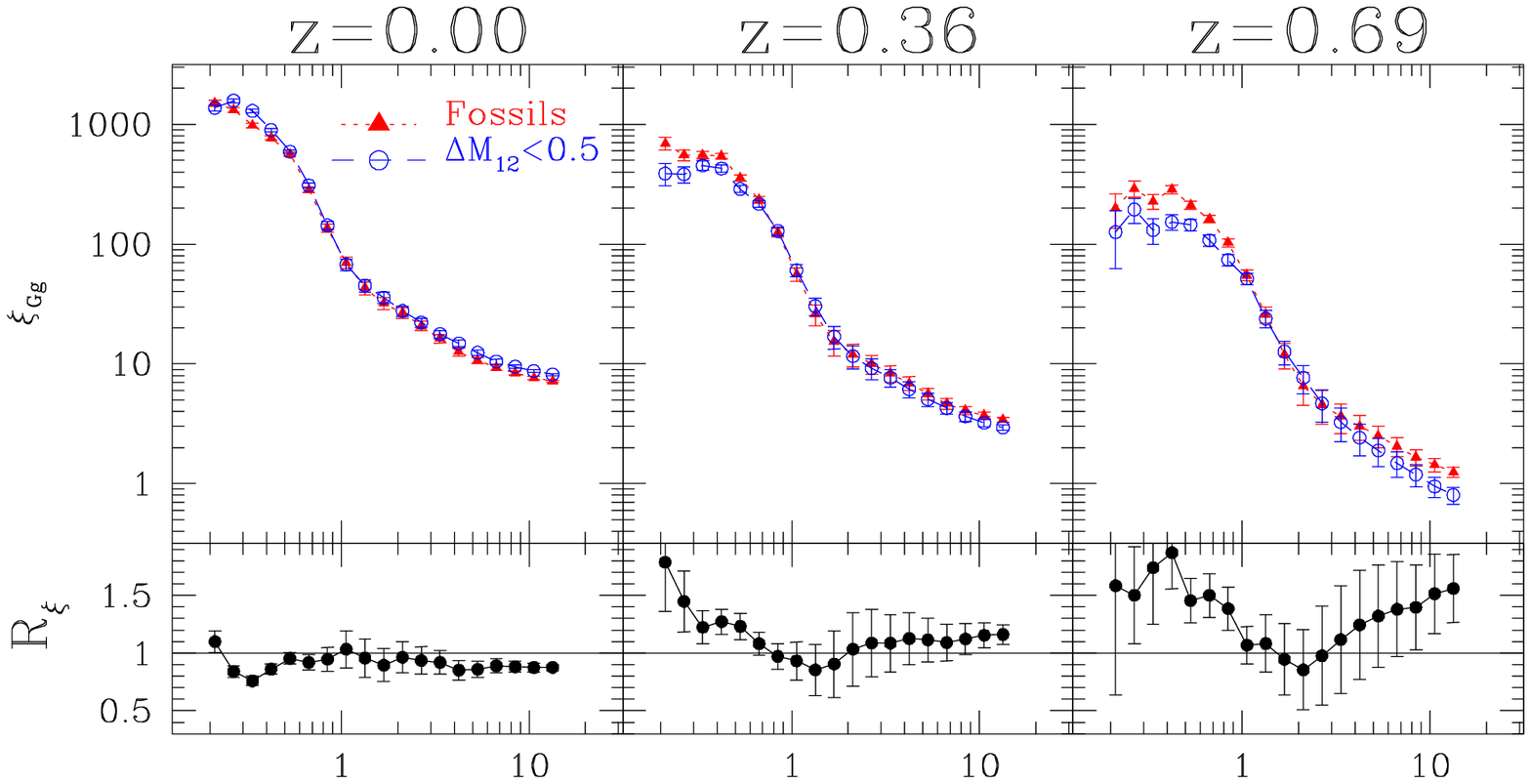}}
{\includegraphics[width=15cm]{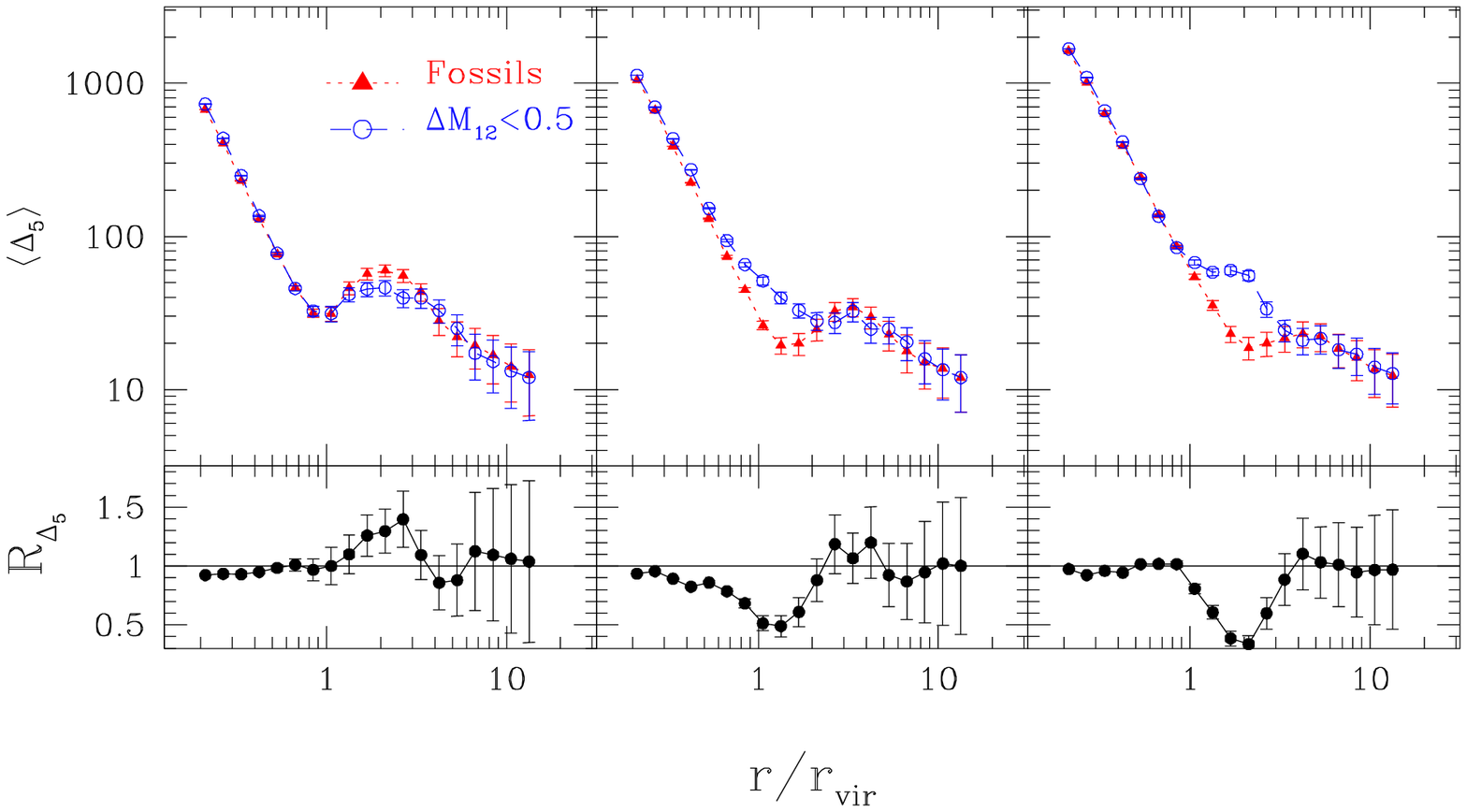}}
\caption{\emph{Top panels}: Group-galaxy cross-correlation functions
  of the mass-matched samples ($\xi_{G-gal}$). \emph{Upper panels}:
$\xi_{G-gal}$ in the MS as a function
  of the normalised group-centric distance.  \emph{Filled triangles}
  (\emph{dotted lines}) correspond to fossil groups, while \emph{open
    circles} (\emph{dashed lines}) correspond to normal groups with
  a magnitude gap $\Delta M_{12}$ smaller than $0.5 mag$.  Error bars are
  computed using an analytic formula for the Poisson error in the
  clustering signal \citep{Peacock99}.  \emph{Lower panels}: Ratios
  of cross-correlation functions ($R_{\xi}$) for fossils and groups
  with $\Delta M_{12}<0.5 mag$. Error bars are computed using the
  usual formula of error propagation.  \emph{Panels from left to
    right}: Different stages of the group evolution (see upper
  labels). 
\emph{Bottom panels:} Local over-density profile of galaxies
  around groups ($\Delta_{5}$). \emph{Upper panels}: 
$\Delta_{5}$ in the MS as a function of the
  normalised group-centric distance.  \emph{Filled triangles}
  (\emph{dotted lines}) correspond to galaxies around fossils groups,
  while \emph{open circles} (\emph{dashed lines}) correspond to
  galaxies around normal groups with $\Delta M_{12}<0.5~mag$.  Error bars
  represent the 35th and the 65th percentiles in the local density
  distribution at each distance bin.  \emph{Lower panels}: Ratios
  of local density profiles ($R_{\Delta_{5}}$) of fossils and
  groups with $\Delta M_{12}<0.5 mag$. Error bars are computed using the
  usual formula of error propagation.  \emph{Panels from left to
  right} correspond to the stages of the group evolution (see upper labels).
% i.e., 
%following groups back in time at different snapshots in the MS.
}
\label{xi}
\end{figure*}

\section{The samples}
\label{samples}
Our samples of fossil and normal groups were extracted from the
Millennium Simulation \citep{Springel+05} combined with a semi-analytic model of galaxy
formation. These samples were defined in \cite{diaz08}. Here, we
briefly summarise their main characteristics:

\begin{itemize}

\item The MS evolved 10 billion ($2160^3$) dark matter particles
  within a periodic box of $500 \ h^{-1} \ Mpc$ in a Lambda cold dark
  matter ($\Lambda CDM$) cosmology with parameters: $\Omega_m=0.25$,
  $\Omega_{\Lambda}=0.75$, $\sigma_8=0.9$, and $h=0.73$.

\item To extract galaxies from the MS, we adopted the run of
  \cite{dLB07}'s semi-analytic model, which provides positions,
  velocities, absolute magnitudes (BVRIK), etc. The final output at
  $z=0$ contained $\sim 10^7$ galaxies with absolute
  magnitudes $M_R -5 log \ h < -17.4$ and stellar masses higher than
  $3\times10^8 \ h^{-1} \ {\cal M}_{\odot}$.

\item Groups were identified at the z=0 output of the simulation
  using a friends-of-friends algorithm in real space with a linking
  length of 0.2 of the mean particle density and cross-correlated with the
  DM halos identified in the MS (\cite{diaz08}).  Their merging
  history was followed back in time to study the evolutionary behaviour
  of galaxies inside and outside the virial radii of the
  systems. Virial radii and positions of the groups at the different
  outputs used throughout this work were extracted from the publicly
  available information of the MS+semi-analytic model\footnote{The
  Millennium simulation, performed by the Virgo Consortium, is available
  at http://www.mpa-garching.mpg.de/millennium}. Specifically, virial
  radii were estimated using the virial theorem, with the virial masses
  and velocity dispersions obtained from the MS database. The virial
  masses used for these estimates were the masses within the radius
  where the halo has an over-density corresponding to the value at
  virialisation in the top-hat collapse model for the $\Lambda CDM$
  cosmology.
 
\item Fossil systems were selected from the sample of systems
  identified at $z=0$ to have virial masses higher than $5\times10^{13}
  \ h^{-1} \ {\cal M}_{\odot}$ and absolute magnitude difference
  between the first and second brightest galaxies ($\Delta M_{12}$)
  greater than 2 (in the R-band) when considering objects within $0.5
  \ r_{vir}$. This lower cut-off in virial masses maximises the
  probability that the selected systems are X-ray emitters
  \citep{Dariush07}. The final sample comprises 591 fossils. We note
  that there is a small difference between this sample and the one
  obtained by \cite{diaz08}. This is due to the lack of some physical
  properties of some systems in particular redshift snapshots used in
  our analysis.  The virial mass distribution of fossil systems can be
  seen in the \emph{upper left panel} of Fig.~\ref{masmatch}
  (\emph{empty histogram}).

 \item Following the procedure of \cite{diaz08}, we define a sample of
   non-fossils in order to directly compare with the fossil
   sample. The non-fossil sample comprises groups with the same lower
   limit in virial mass, but having $\Delta M_{12}<0.5 mag$. This
   sample of non-fossils comprises 1997 systems. It is well-known that
   cluster formation histories depend strongly on the mass of the
   systems. Consequently, differences in the virial mass
   distributions of the fossil and non-fossil samples could
   introduce bias into the results. Therefore, we extracted a
   sub-sample of 591 non-fossil groups with the same virial mass
   distribution (Kolmogorov-Smirnov coefficient $>$ 0.98) as the
   sample of fossils (the ``mass-matched'' samples; see \emph{upper
     left panel} of Fig.~\ref{masmatch}).  The \emph{lower left panel}
   of Fig.~\ref{masmatch} shows the assembly time distributions (time
   at which groups have assembled half of their final virial mass) for
   the samples of fossil and non-fossil ($\Delta M_{12}<0.5~mag$) groups.

\item To test the effects of assembly time on our statistics,
  we also defined samples of fossils and non-fossils with matched
  assembly times (right panels, Fig~\ref{masmatch}). We refer to
  these as the ``assembly-matched'' samples (although
  we note that the samples are both assembly-time- \emph{and}
  mass-matched).

\item We also studied the remaining sample of normal groups with
  $0.5\le \Delta M_{12}\le 2$, which comprises 6881 groups. We divided
  this sample of normal systems into 15 sub-samples, each of them
  resembling the fossil sample in terms of the number of systems and virial mass
  distributions.
\end{itemize}

  We note that we are unable to easily assess the effects of using
  a semi-analytic model other than the \cite{dLB07} model used
  in this work. However, given that all similar models that reproduce the
  galaxy luminosity function reasonably well, we would expect any
  effects to be small. We particularly note that the \citeauthor{dLB07} model is one
  of the best at reproducing the bright end of the luminosity function
  \citep{BDLT07}, which is of particular importance when measuring the
  $\Delta M_{12}$ parameter that defines fossil groups.

  We note that our fossil sample is comprised of
  groups that exhibit $\Delta M_{12}> 2$ at z=0, and does not include
  groups that had $\Delta M_{12}> 2$ at higher redshift, but became
  'normal' again by z=0 (as suggested by \citealp{vonbenda08}).  The
  presence of these groups in our normal group samples can only serve
  to reduce any differences between our fossil and non-fossil samples
  and therefore does not affect our {\it qualitative} analysis.

\section{The group-galaxy cross-correlation function}
\label{xi_cg}
The spatial group-galaxy cross correlation function, $\xi_{Gg}(R)$, is
defined such that the probability $dP$ of finding a galaxy in volume
element $dV$ at distance $R$ from the centre of a group is
\[dP=\bar{\eta} \ [1+\xi_{Gg}(R)]dV,\]
where $\bar{\eta}$ is the mean number density of galaxies in the whole
MS. The quantity $\xi_{Gg}(R)$ is therefore equivalent to the radially averaged
number over-density profile. To increase the statistical
significance of our results, we averaged the values for all groups
after scaling to take into account the different group sizes. This is
achieved by normalising the group-centric distances to the virial
radius of each group ($R=r/r_{vir}$). We refer to everything
that lies farther than one virial radius from the group centre as
the ``environment'' of the system.

At each stage in our analysis, a comparison was made between normal
groups having $0.5 mag<\Delta M_{12}<2 mag$ and those with $\Delta
M_{12}<0.5 mag$. We found no differences between these two samples of
normal groups at any stage.  Therefore, throughout this paper we
present the comparison of our fossil sample with the $\Delta
M_{12}<0.5 mag $ sample.

To follow the evolution of the environment around groups, we
chose three different redshift snapshots in the MS: z= 0.00,
0.36, and 0.69. The highest redshift value (0.69) represents the time
by which $90\%$ of fossil groups have assembled half of
their final virial mass, while the middle redshift value (0.36)
represents the time by which $90\%$ of non-fossil groups have
assembled half of their final virial mass. These characteristic times 
are inferred from the assembly time distributions shown in the 
lower left panel of Fig.~\ref{masmatch}.

The top panels of Fig.~\ref{xi} show the cross-correlation 
functions obtained for the
mass-matched samples of fossil and non-fossil ($\Delta M_{12}<0.5
mag$) groups, where the evolution of these functions is shown by the three
redshift snapshots. We explored the cross-correlation functions out to
a normalised distance of $\sim 10 \ r_{vir}$ which represents a median
distance of $\sim 17 \ {\rm h^{-1} Mpc}$
(see. Sect.~\ref{delta}). Since the mean-intergroup separation for
systems with virial masses higher than $5\times10^{13} \ h^{-1}\,
{\cal M}_{\odot}$ is approximately $20 \ {\rm h^{-1} Mpc}$
\citep{ZMP03}, this limit in normalised clustercentric distances
minimises the risk of the spatial superimposition of groups in our main
samples. At all stages, all these cross-correlation functions show a
clear transition from the `one-halo' to the `two-halo' regimes at the
virial radius of the groups \citep{yang05}. In general, the one-halo term dominates the
total contribution in the non-linear regime, while the two-halo term
captures the large-scale correlations in the linear regime.

Starting at $z=0.69$ (Fig.~\ref{xi}; upper right panel), the number
density within the groups ($<1~r_{vir}$) is higher for fossils than
non-fossils. At this redshift, this may be because most fossil groups unlike
normal groups, have already assembled half of their
final masses.  Beyond 3~$r_{vir}$, the environments of fossils exhibit higher
densities than normal groups. This is consistent with the earlier
formation time of fossil groups, since halos forming in higher density
environments collapse earlier \citep{sheth04,Harker06}.  
There is a transition region between 1 $r_{vir}$ and 3 $r_{vir}$ where the
environments around fossils and non-fossils looks similar (in terms
of number density).
 
By z=0.36, both fossil and non-fossil systems have clearly
evolved. The evolution both inside and outside of the virial radius is
such that fossil and non-fossil systems have become more
similar. Within the virial radius, this is to be expected since, by
this time, most normal groups have also assembled more than 50\% of
their final mass.

At z=0, the ratio $R_{\xi}=\xi_{Fg}/\xi_{Ng}$ indicates that fossils
appear to have slightly fewer galaxies within half a virial radius than
normal groups.  This may simply reflect that the central
galaxies of fossils have cannibalised their neighbours to produce the
two magnitude gap.  There is again a 'transient' region between 0.5
$r_{vir}$ and 2.5 $r_{vir}$ where the global densities around both
classes look very similar.  Beyond this region, the environment of
fossil systems at redshift zero has a slightly \emph{lower} density
than around normal systems.

\begin{figure}
%  masa.sm --> rvir_z
\centering
{\includegraphics[width=8cm]{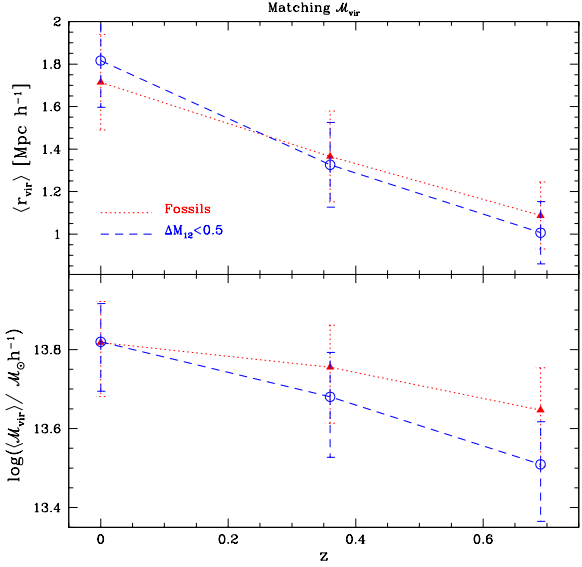}}
\caption{Median values of virial radii (\emph{upper panel}) 
and masses (\emph{lower panel}) as a function of redshift for fossils
  (\emph{dotted line}) and $\Delta M_{12}<0.5 mag$ groups (\emph{dashed
    line}). Error bars are the 25th and 75th percentiles. 
   }
\label{rvir_z}
\end{figure}

A possible cause of the local density differences at high
redshift is revealed in the \emph{lower panel} of
Fig.~\ref{rvir_z}, which shows the evolution of the halo masses with
redshift.  Although both fossil and non-fossil samples reach the
same final masses (by construction), at earlier times fossil groups
can be seen to have higher masses. This figure therefore
suggests that the mass growth rate of fossils differs from that
of non-fossils. However, these differences in halo mass evolution
could be a consequence of different assembly times, since halos
forming earlier are thought to form in higher density regions and
also to have had more time to gain mass. This point will be addressed in
Sect.~\ref{differences}, where we discuss the mass- \emph{and}
assembly-time matched samples.

 \begin{figure}
% den_za.sm --> meds2f
\centering
{\includegraphics[width=8cm]{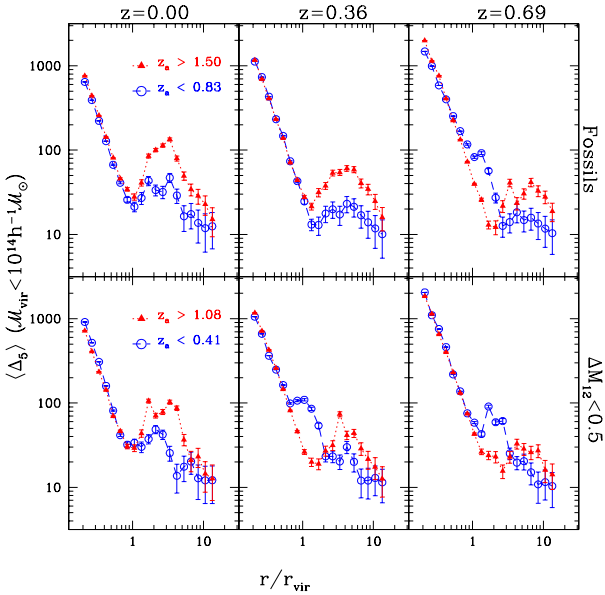}}
\caption{Local over-density profile of galaxies around fossil
  (\emph{upper panel}) and non-fossil (\emph{lower panel}) groups in
  the MS as a function of the normalised group-centric distance.
  \emph{Filled triangles} correspond to the sub-sample of systems with
  early assembly times, while \emph{open circles} correspond to a
  sub-sample with later assembly times.  Error bars are the 35th and the
  65th percentiles of the local density distribution at each distance
  bin.  }
\label{denza}
\end{figure}

%%%%%%%%%%%%%%%%%%%%%%%%%%%%%%%%%%%%%%%%%%%%%%%%%%%%%%%%%%%%%%%%%%%%%%%%%%%

\section{The local density profile}
\label{delta}
%%%%%%%%%%%%%%%%%%%%%%%%%%%%%%%%%%%%%%%%%%%%%%%%%%%%%%%%%%%%%%%%%%%%%%%%%%%%
To continue our analysis of the environment around galaxy systems, we
also study the local over-density profile $\langle \Delta(r/r_{vir})
\rangle$. This statistic measures the local density around each galaxy
and therefore contains information about the local clustering (clumpiness) of
the groups.

The local density, $\eta_N$, is measured for every galaxy in and around
groups. It is computed as the number density within a sphere of radius
defined by the distance ($d$) to its {\it N-th} nearest neighbour, i.e.\[
\eta_N=\frac{3 N}{4\pi d_N^3}.\] 
The local over-density is then given by
\[ \Delta_N = \frac{\eta_{N}- \bar{\eta}_N}{\bar{\eta}_N}, \]
where $\bar{\eta}_N$ is the mean local density in the simulation.  The
local over-density profile is obtained by computing the median values
of the over-density associated with each galaxy as a function of the
3D-normalised cluster-centric distance, i.e., $\langle \Delta_N (r/r_{vir}) \rangle$. The
$N$ value was tested using two different values of 5 and 10, and
very similar results were obtained. We therefore show all our results using $N=5$.
As in the previous section, no significant differences were found
between the results for normal groups having $0.5~mag<\Delta M_{12}<2~mag$ and
those with $\Delta M_{12}<0.5~mag$.

The local over-density profile is shown in the bottom panels of
Fig.~\ref{xi}. It can be seen that, as expected, the local
over-density of galaxies diminishes as we approach and pass through
the virial radius. However, there is an increase in the local
over-density around 2-4 $r_{vir}$ (the "bump")\footnote{We refer
to the region where the local density increases outside the virial
radius as the "bump".}. This suggests that either there is a sudden
increase in the global density at this radius, \emph{or} 
that galaxies at these distances are more clustered than their
surroundings. The first hypothesis was ruled out by our study of the
cross-correlation function in the previous section.
We note that \emph{both fossils and non-fossils show this
increase in all redshift snapshots}, although it is more pronounced
for regions around fossil groups at $z=0$.

\begin{figure*}
%  den_eq.sm --> 
\centering
{\includegraphics[width=15cm]{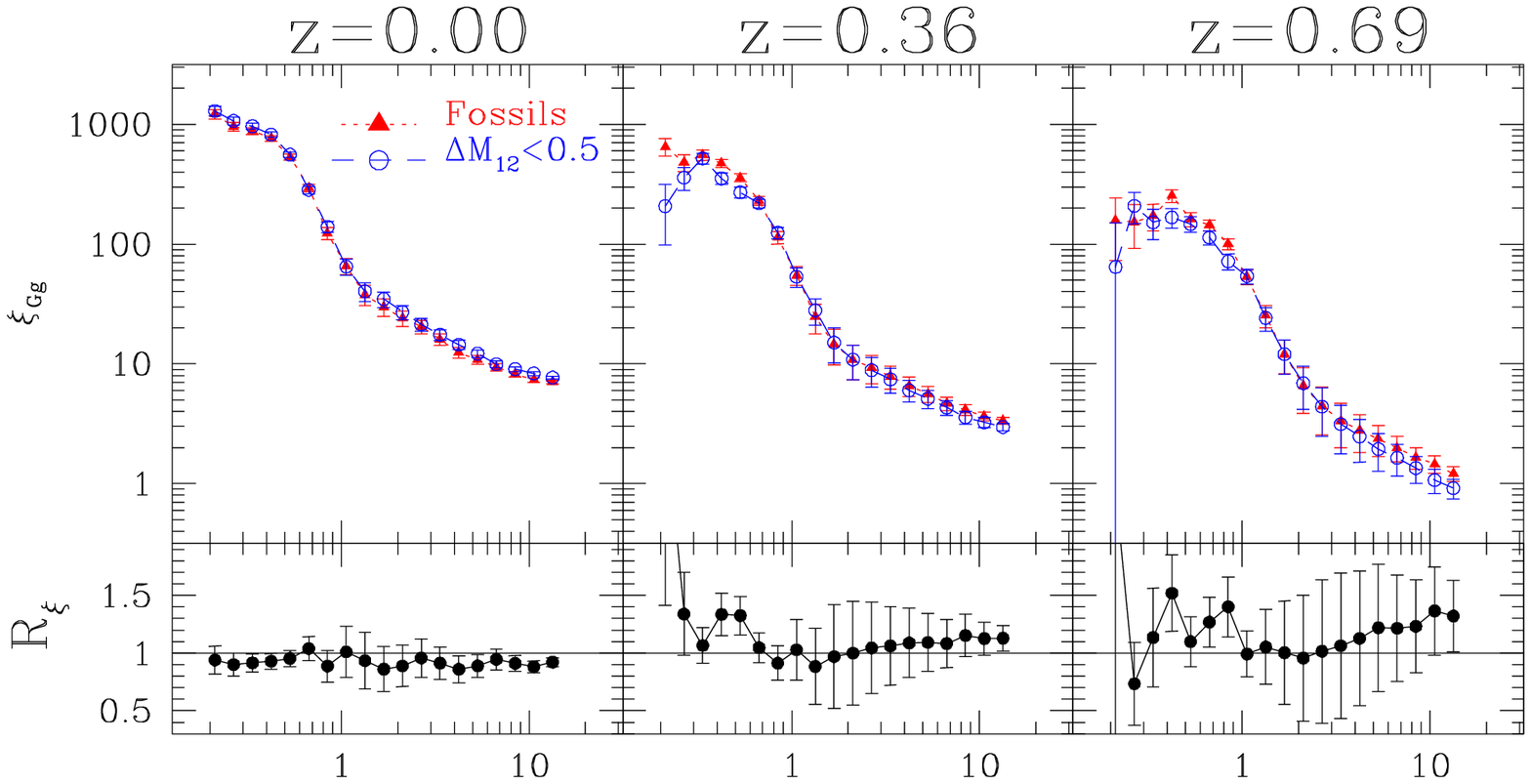}}
{\includegraphics[width=15cm]{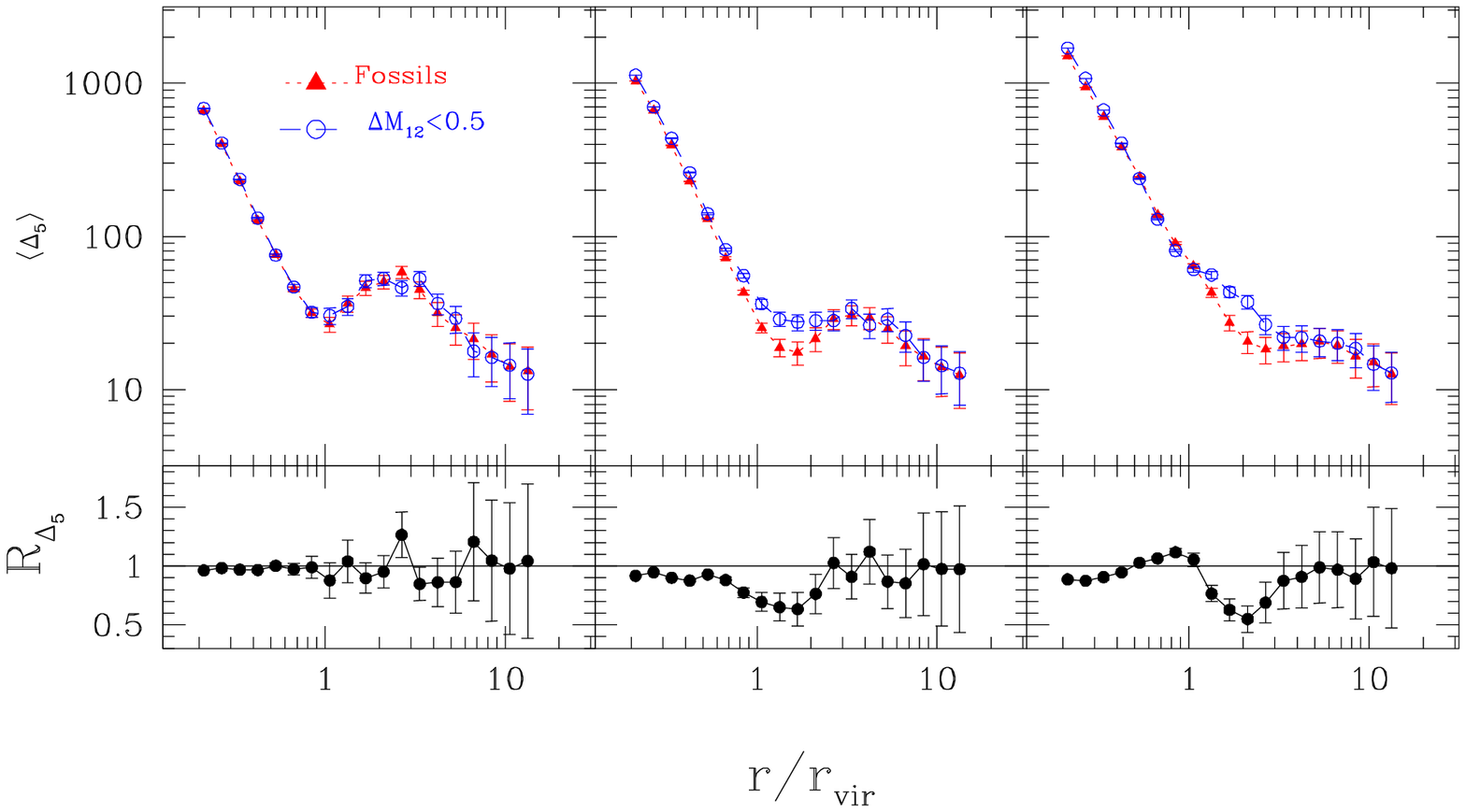}}
\caption{Same as Fig.~\ref{xi} but for the sub-samples of fossils and
  non-fossils with the same virial mass and assembly time
  distributions (samples in right panels of Fig.\ref{masmatch}). }
\label{den_eq}
\end{figure*}

\subsection{The behaviour of the bumps}
In this section, we analyse the position and amplitude of the bump
observed in the local density profile, focusing on their depedence on 
group properties.

\subsubsection{The apparent movement}
When analysing the bumps for the different evolutionary stages, it can be
seen that they become more pronounced towards z=0, but that they also seem
to be moving closer to the group. Our analysis shows that most of the
galaxies located at the position of the bump at a given redshift, are
also positioned in the region of the bumps in later outputs. This
finding leads us to rule out the possibility of rebound galaxies
\citep{Mamon04,Gill05,Mamon06_Chile} being the cause of the bump since
these galaxies are expected to dramatically change their positions
with respect to the group centre between redshifts 0.69 and 0.00. The
apparent movement of the bump could therefore be the result of either
infalling substructure \emph{or} simply the growth of the virial
radii of the systems concerned, i.e. the physical location of the bump
being fixed. The \emph{upper panel} of Fig.~\ref{rvir_z} shows the
variation in the virial radius with time. It can be seen that the
median of the virial radius for fossils increases from $\sim$1.1$\ Mpc
\ h^{-1}$ at $z=0.69$ to $\sim$1.7$\ Mpc \ h^{-1}$at $z=0.00$.
The bump in the local density profile for fossil groups at z=0.69 is
positioned at $r/r_{vir}\sim 4.5$, which in physical units corresponds
to $\sim 5 \ h^{-1} Mpc$. If the galaxies in the bump possess no
significant proper motions with respect to the cluster centre, then we
should find the bump at 2.9$\ r_{vir}$ at redshift z=0.00, i.e., we
should see an \emph{apparent} movement simply because the size of the
groups increases with time. However, it can be seen that the position of
the bump at z=0.00 is $\sim 2.5 \ r_{vir}$, which indicates that
besides the apparent movement caused by the change in size, there is
also a real movement of the particles in the bump towards the centre
of the group. The proper motions of galaxies in the bumps have to
account for a movement of less than $0.5 r_{vir}$, which means that those galaxies
have to have radial velocities with respect to the centre of the groups of
$0.5 r_{vir}/\Delta t \sim 850 h^{-1} kpc/ 6.118Gyr=1164 kpc /
6.118Gyr \sim 185 \ km/s$.

\subsubsection{Dependence on the assembly times}
It is well known that both the mass and assembly times of the systems
are affected by their environment \citep{espino07}. It is widely
accepted that fossil systems assembled earlier than non-fossils
\citep{Donghia,Dariush07,diaz08}.  Hence, the following question arises: how does
the assembly time of the systems affect the local density profiles?
To answer this question, we re-sampled the mass-matched samples
according to their assembly times\footnote{According to the definition of
\cite{delucia06}, we take the assembly time $z_a$ to be the time when the
system has acquired $50\%$ of its final ($z=0.0$) virial mass.}.
We also restricted the samples
to ${\cal M}_{\rm vir}<10^{14} {\rm h}^{-1} {\cal M}_\odot$ to avoid
differences being caused by any differences in mass. 
Figure~\ref{denza} shows the local density
profiles for systems with early assembly times ($>80$th percentile of the
distribution of assembly times)(\emph{dotted lines}), and systems
that assembled later ($<20$th percentile of $z_a$) (\emph{dashed lines}).  

The effect of the assembly time on the local density profile is clear:
\emph{Systems that assembled earlier show a more pronounced bump 
than systems that assembled later}, and this is true for both fossils
\emph{and} non-fossils.

It can also be seen that the local density profiles of non-fossil
systems that assembled later ($z_a<0.41$) show an inner peak, 
very close to the virial radius for the snapshots z=0.36 and z=0.69 
(\emph{middle and right lower panels}), 
and this feature can also be seen in the
local density profile at z=0.69 for fossil groups with late assembly 
times ($z_a<0.83$) (\emph{right upper panel}).
The snapshots at z=0.36 and 0.69 represent stages where the groups
with late assembly time are not yet fully
assembled (in the sense that they have not reached half of their final
mass) and are therefore less relaxed. 
On the other hand, groups
(fossils or non-fossils) with early assembly times do not show any
inner peak immediately outside the virial radius. 
Their earlier assembly makes these
groups more relaxed in the snapshots shown in
Fig.~\ref{denza}. Therefore, the shape of the over-density profile is
then influenced by the assembly time of the groups, and also by the
stage of relaxation of the groups in any given snapshot. 

\section{Differences between fossils and non-fossils}
\label{differences}
We have demonstrated above that the local density profiles of fully
assembled galaxy systems increase beyond the virial radius. We
have found that galaxies remain in this region for a long time,
gradually approaching the group centres. We have also shown that
the shape of the local density profile and the amplitude of the bumps
strongly depend on both the assembly times of the groups and
their relaxation stage. This last finding might lead us to conclude
that the environmental differences observed between fossil and
non-fossil systems are purely an evolutionary effect since fossil
groups assembled earlier. However, although fossil groups are indeed,
on average, older than non-fossils, there exist sub-samples of fossils
and non-fossils that assembled at similar times (see \emph{lower left
panel} of Fig.~\ref{masmatch}).  Since fossils have developed the
two-magnitude gap, while non-fossils do not, we are led to ask: Could
the environment be responsible for this magnitude
gap in fossils?

In this section, we therefore investigate whether the environment
is different for fossil and non-fossils systems that have the same
assembly times, and consequently, the same relaxation in any given
snapshot.  To this end, we select new sub-samples of fossils and
non-fossils having the same z=0 virial mass \emph{and} assembly time
distributions (the assembly-matched samples; \emph{upper right} and
\emph{lower right} panels of Fig.~\ref{masmatch}). Both of the new samples
comprise $\sim 330$ groups each.\\

\begin{figure}
%  masa.sm --> radio_masa_z
\centering
{\includegraphics[width=8cm]{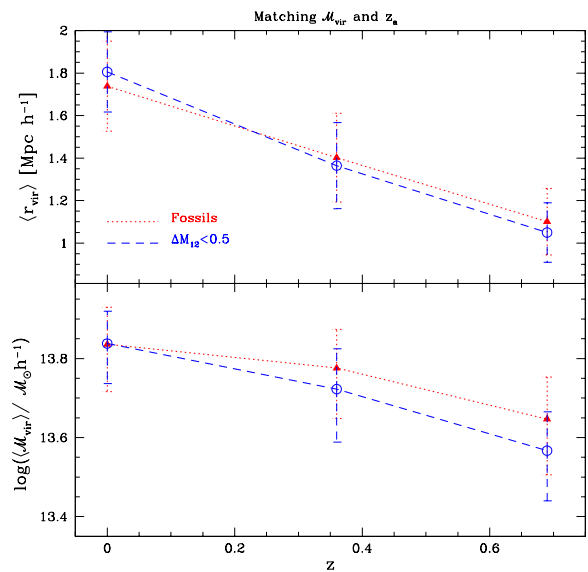}}
\caption{Median values of virial radii (\emph{upper panel}) 
and masses (\emph{lower panel}) as a function of redshift for fossils
  (\emph{dotted line}) and $\Delta M_{12}<0.5 mag$ groups (\emph{dashed
    line}) with the same mass and assembly time distributions.
  Error bars are the 25th and 75th percentiles.  }
\label{rm_mz}
\end{figure}

We first calculate the cross-correlation function in order to check
whether the differences observed in the \emph{upper panels} of
Fig.~\ref{xi} persist. The upper panels of Fig.~\ref{den_eq} show the
cross-correlation function for these new samples. Comparing these
figures, the strong differences seen, particularly at the earliest
snapshot (z=0.69), have become smaller. However, the tendency of fossil
groups to be in higher number density regions at high z is still
present, although it is clearly somewhat weaker. To confirm
this trend, we analysed the evolution of the halo virial masses for
the assembly-matched samples. In the \emph{lower panel} of
Fig.~\ref{rm_mz}, the medians of the halo masses at the different
snapshots are shown. As can be seen, at z=0, the fossil and
non-fossil halos have the same mass (by construction). However, at
higher z, fossils exhibit higher halo masses.  This is an interesting
result since both samples have the same final mass and assembled at
the same time. The result supports the weak trend observed at high z
in the cross-correlation function. A higher number density around
fossils clearly allows these groups to reach higher masses than non-fossils in
the same period of time. Hence, the environment of fossils and non
fossils are different even for the assembly-matched samples,
indicating that the differences are a matter not only of time
evolution.

We next calculate the local density profiles for the assembly-matched
samples. The \emph{bottom panels} of Fig.~\ref{den_eq} show the local
over-density profiles for these new samples.  Comparing these
results with those found in Sect.~\ref{delta}, it can be seen that
the local over-density inner peak observed just outside the virial radius
of non-fossils at z=0.36 and 0.69 is drastically
reduced, as a result of our having discarded the least relaxed
non-fossil systems. 
Moreover, at z=0, the bumps for fossils and non-fossils
became similar, which indicates that the difference observed 
for the mass-matched samples at this snapshot is due only to differences
in the assembly times of the groups involved.
However, \emph{the main differences between 
fossils and non-fossils observed at earlier evolutionary stages are still present despite
there no longer being any assembly time dependence}. This is a
very interesting result as it indicates that the environments of
fossils and non-fossils are indeed intrinsically different at high redshift, 
and may therefore be the cause of the differences in the magnitude gaps
among these systems.

\section{Comparison with observations}
\label{observations}
%%%%%%%%%%%%%%%%%%%%%%%%%%%%%%%%%%%%%%%%

The most interesting result we have found is related to the local
over-density profile at redshift zero. From this profile, we observe
that the environments of \emph{simulated} fossil systems look different
from the environments of normal groups.  Is it possible to observe the
same behaviour of the local over-density for a sample of observational
fossil systems?  

To answer this question, we selected small samples of fossil and normal
systems from a large-scale galaxy survey.  We used the main
spectroscopic sample of galaxies of Sloan Digital Sky Survey Data
Release 7 (SDSS DR7, \cite{DR7}) and, following the work of
\cite{heitmann09}, selected four known fossil groups from their
sample. We also constructed a sample of four normal groups from the 400d
Cluster Catalogue \citep{burenin07}. The normal groups were selected
to ensure that the groups had similar radial velocity dispersions and mean
redshifts to those in the fossil sample. The velocity dispersion
restriction was imposed in order to match the virial mass distribution
criterion adopted in the simulations, while the redshift restriction
was designed to reduce the number density dependence on distance.
The virial radii of the systems were computed from their X-ray
luminosity following \cite{Vik09} and \cite{heitmann09}. Details of
the systems are given in Table~\ref{observable}.

In the work presented above, we have defined the local over-density in the
simulation using the full three-dimensional (3D) spatial information. This is obviously
not possible with observational data. In addition, the observable samples of
galaxies are apparent-magnitude-limited, and the computed local
densities will therefore depend on the redshifts of the
groups. Consequently, we slightly modified the way of measuring
local over-densities in the observational data. Our procedure is as
follows:
\begin{enumerate}
 \item We select a spectroscopic sample of galaxies around each group consisting of
   galaxies within $1000 \ km/s$ of the group centre to avoid projection effects.
 \item We then select galaxies brighter than the absolute magnitude limit that
   corresponds to the farthest redshift allowed ($v_{max}=v_r + 1000
   \ km/s$) (volume limited sample).
 \item We measure {\it projected} local densities using the third closest
   neighbour ($\Sigma_3$).
 \item We compute the mean projected local density corresponding to
   the redshift of the group $\bar \Sigma_3(z)$.
 \item We compute the projected local over-density as
   $\Delta_3=\frac{\Sigma_3-\bar \Sigma_3(z)}{\bar \Sigma_3(z)}$.
 \item We compute the projected local over-density profile as a function
   of the normalised {\it projected} distance to the centre.
\end{enumerate}

\begin{table}
 \begin{center}
% use packages: array
\caption{Observational Fossil and Normal Groups \label{observable}}
\begin{tabular}[c]{|l|c|c|c|c|}
\hline
\hline
Fossil Groups & z & $\sigma_v$ & $r_{\rm vir}$  & $L_X$ \\ 
 & × & [$km \ s^{-1}$] & [$h^{-1} \ Mpc$]  & [$erg \ s^{-1} \ 10^{43}$] \\ 
\hline
RXJ1340+4018 & 0.171 & 706 & 1.05 & 1.24 \\ 
RXJ1159+5531 & 0.081 & 435 & 1.08 & 1.14 \\ 
RXJ1416+2315 & 0.138 & 500 & 1.49 & 6.09 \\ 
RXJ1552+2013 & 0.136 & 459 & 1.22 & 2.29 \\ 
\hline
Normal Groups & × & × & × & × \\ 
\hline
400d-177 & 0.169 & 680 & 1.22 & 2.55 \\ 
400d-081 & 0.064 & 449 & 1.25 & 2.24 \\ 
400d-151 & 0.160 & 502 & 1.06 & 1.26 \\ 
400d-180 & 0.167 & 452 & 1.33 & 3.88\\
\hline
\hline
\end{tabular}
\parbox{\hsize}{Notes: Normal groups were extracted from the 400d
Cluster Catalogue http://hea-www.harvard.edu/400d/catalog/table\_cat.html}
\end{center}
\end{table}

\begin{figure*}
% observable/den.sm --> ?
\centering
{\includegraphics[width=8cm]{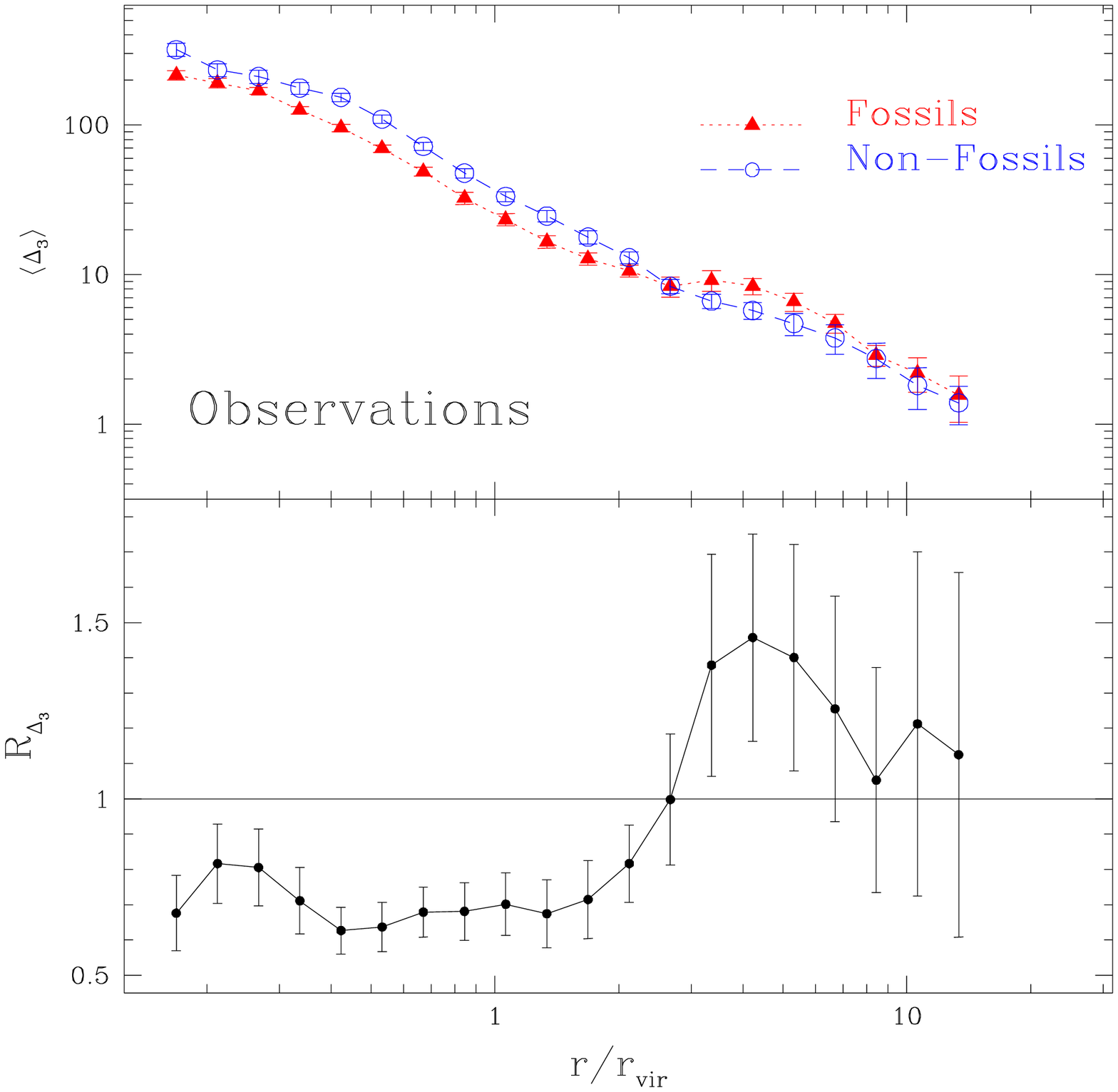}
\includegraphics[width=8cm]{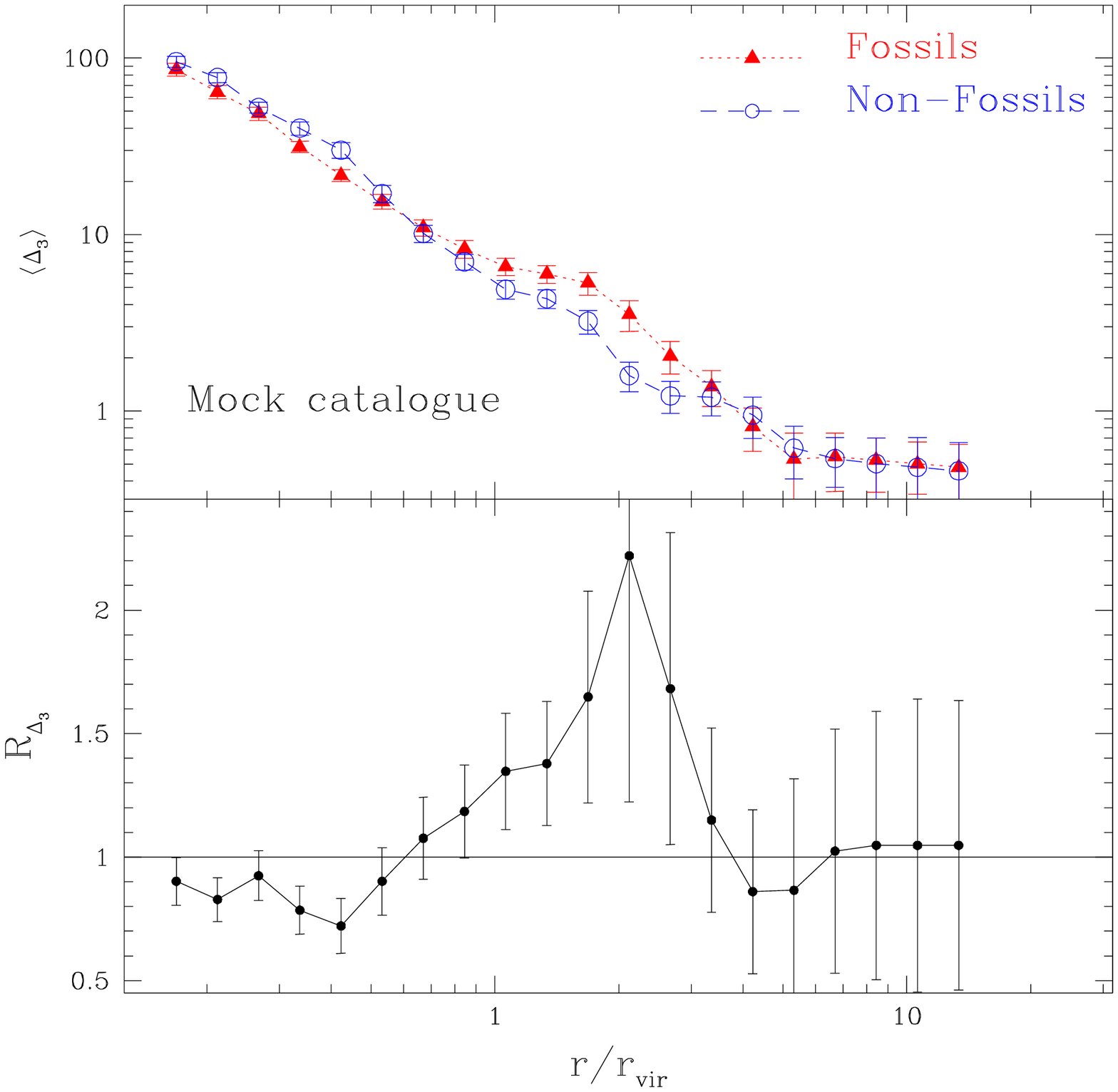}}
\caption{Local density profile of galaxies around groups as a function
  of the normalised group-centric distance. Left upper panel show the
  local density profiles for fossils and normal systems identified
  observationally (see Table 1) while right upper panel show the
  profiles for Fossils and $\Delta M_{12}<0.5 mag$ groups in the MS when
  introducing a brighter cut-off in the R-band absolute magnitude.
  Error bars are the 35th and the 65th percentiles in the local density
  distribution at each distance bin. Lower panels show the
  corresponding ratios of local density profiles in each case.}
\label{den_obs}
\end{figure*}

The projected local over-density profiles in the observational data
are shown in the left panel of Fig.~\ref{den_obs}. This figure
shows that the bump observed beyond the virial radius in the
simulated fossil groups is also present in the observational sample.
However, it can also be seen that the ratio of the local
over-density profiles (R$_{\Delta_5}$) \emph{within} 1 $r_{vir}$ of
systems in the observations differs from that shown in Fig.~\ref{xi}
(lower left panel), with galaxies in fossil systems exhibiting
significantly lower relative local over-densities than their
counterparts in normal systems. However, this can be understood in
terms of the differing luminosity limits applied to the two
analyses, i.e., since we consider fossils, we expect, by
definition, to find a lack of bright galaxies in the proximity of
the central galaxy compared to similar regions in the normal
systems, while faint galaxies should be equally distributed in both
samples.  This is only discernible when working on
observational samples that are magnitude limited, since only the
brightest galaxies are observed.

To perform a fair comparison between observations and semi-analytic
models, it is particularly helpful to construct a mock galaxy catalogue. 
We therefore used the semi-analytical model to build a mock sample of
galaxies within a cone consisting of shells constructed from different
snapshots corresponding to the epoch of the look-back time at their
distance. Here we use the 17 last snapshots, bringing us to a maximum
redshift of $z=0.68$. For the mock galaxies, we obtain redshifts by
adding the Hubble flow to the peculiar velocities projected along the
line-of-sight direction. We compute the observer-frame galaxy
apparent magnitudes from the rest-frame absolute magnitudes provided
by the semi-analytical model. These apparent magnitudes are converted
to the observer frame using tabulated $k+e$ corrections
\citep{Poggianti97}. We set an apparent magnitude limit $R=17.44$,
which roughly corresponds to the r-sdss band magnitude limit of
$r=17.77$ (see Appendix A in \cite{diaz10}).  We identify groups of
galaxies in the mock catalogue following a similar procedure as
\cite{z06}. We also measure virial masses, virial radius, and radial
velocity dispersions for the groups following the cited work.

Fossil and non-fossil groups are selected according to the criteria
described in Sect.~\ref{samples}.  We found ten fossil groups in the
mock catalogue with a median redshift of $0.064\pm0.015$ and median
group radial velocity dispersion of $338\pm78 \, km \, s^{-1}$.  We
selected ten non-fossil groups among the $219$ identified in the mock
catalogue to reproduce the group radial velocity dispersion
and redshift distributions of the fossil groups. The local
over-density profile is then measured by using the same procedure
described for the observational sample of groups and is shown in the
\emph{right panel} of Fig.~\ref{den_obs}. Comparing both panels, it
can be seen that there is roughly an overall agreement between the
profiles computed from the observations and the mock galaxy
catalogue. The only noticeable difference is that the position of the
bump in the observations is shifted towards larger cluster-centric
distances.  This shifting might be, in part, a consequence of
analysing slightly different evolutionary stages (z=0.13 vs. z=0.06),
which would lead to the position of the bump being altered. However, this
difference is also affected by the different methods employed to
evaluate the virial radii (which are based on values estimated from
X-ray observations for the SDSS observational sample, and the
virial theorem radii in the mock catalogue).

\section{Summary}
\label{conclusions}
We have analysed the environments around
fossil and normal galaxy systems identified in the Millennium
simulation \citep{Springel+05} combined with a semi-analytical model of
galaxy formation \citep{dLB07}. Fossil and normal systems have been
selected using the same procedure as \cite{diaz08}. The samples of
normal systems were divided into two sub-samples, taking into account
the magnitude difference between the first and second brightest
galaxies in the systems: $\Delta M_{12}<0.5 mag$ and $0.5 mag \leq
\Delta M_{12}<2 mag$. These sub-samples were constructed to reproduce
the virial mass distribution of fossils to avoid introducing
a mass bias. The results obtained for these two samples of
non-fossils are almost identical. We note
that adopting any specific semi-analytical model introduces a
dependence of the results on the particular set of parameters and
physical processes that were used in the model construction. Since
this work is intended to continue the analyses began by
\cite{diaz08}, analysing the differences caused by
using different semi-analytical models are beyond the scope of
this work. However, we note that the model of \citep{dLB07} used in
this work was shown to reproduce the bright end of the
luminosity function, to which our work is most sensitive, very well
\citep{BDLT07}.

We have investigated the environments using two different approaches,
one global, the other local. We firstly studied the global density of
galaxies as a function of the normalised group-centric distance by
means of the two-point cross-correlation function. Secondly, we
focused on the variation in the local density around galaxies as
a function of the normalised group-centric distance by using the local
density profile. From the global density analysis, we observed that
at z=0.36 and earlier fossil systems in the MS were surrounded by
denser regions than normal systems. This result is consistent with
the common idea that galaxy systems that collapsed earlier lay
preferentially in higher density regions. Following the evolution of
these systems to the present day ($z=0$), we found that the outskirts
of galaxy systems in the MS have evolved differentially, finally, at $z=0$,
reaching a point where the regions surrounding fossils are slightly
\emph{under-dense} compared to normal systems. This last result
agrees with some observational results that studied fossil groups
at low redshifts \citep{Adami07}.\\

From the local density point of view, the local density profiles have
yielded very interesting results. We measured an increase in the
local density profiles in the outskirts of groups for both fossil and
non-fossil samples at $z=0$, but fossils clearly exhibit a sharper 
increase. This local density bump is observed for fossil systems in
every redshift snapshot, decreasing their amplitude and shifting
toward higher distances as we move backwards in time.  The difference
in the local density profiles at $ z=0$ between fossils and
non-fossils occurs at $\sim 2.5 \ r/r_{vir}$, and indicates a clear
difference exists between the environmental distributions.

Even though fossil systems have relatively under-dense regions in their
outskirts according to the global density results, the local density
analysis implies that galaxy clumps are more concentrated around
fossils than around normal systems at z=0.

To address these differences in the environments, we have performed several tests:
\begin{itemize}
\item Firstly, in the local density profile, 
  we have found that the apparent movement of the bump
  with time is not due entirely to the growth of the groups with time, but
  also to the motion of the galaxies in the over-densities towards
  the centres of the groups.
\item Secondly, we divided the samples of both fossil and normal
  systems into two disjoint sub-samples according to their assembly
  times (corresponding to a \emph{early} and \emph{late assembly}). The local density
  profiles demonstrate that, for both samples, the earlier the groups
  assembled, the larger the amplitude in the local density bump is.
  Hence, we have demonstrated that the formation time of groups is
  reflected in the amplitude of this local density feature.  This
  result is quite interesting since the local density profile emerges
  as a new statistical tool suitable for distinguishing the formation
  epoch of galaxy systems at $z=0$, and also defining the virialised
  region of systems.
\item Thirdly, to avoid the assembly-time influence on the
  environment, we have constructed new sub-samples of fossil and
  non-fossil groups with similar virial mass and assembly time
  distributions. The cross-correlation function and 
  the local density profiles for these new sub-samples
  have shown that, even when the groups have formed at similar times, the
  environments around fossils differ intrinsically from those
  observed surrounding non-fossil groups at earlier evolutionary times. 
  Therefore, this result suggests that environment does play an important role in
  the fossil formation scenario.
\end{itemize}

\cite{Porter08} studied structures around clusters, and found
an increment in the star formation activity, peaking at approximately twice the
virial radius of the clusters.  
They hypothesised that an increase in the local density of galaxies at about twice 
the virial radius could explain the increased galaxy star formation in terms of 
galaxy-galaxy harassment.
This theory is strongly supported by the results found in our work here because
we have been able to effectively measure the increase in the local density 
that could lead to efficient galaxy interactions.

Our results also encouraged us to investigate whether these features in
the local density profile are measurable for observational fossil
systems. We have computed the projected local density profile for four
well-known fossil groups and four normal groups, both samples
having similar physical properties. Our results suggest that, even
with poor statistics, we can obtain similar results to those
obtained from the numerical simulations.  We have found a clear
increase in the local density profile for the fossil sample.

\begin{acknowledgements}
We thank the anonymous referee for helpful comments and suggestions.
The Millennium Simulation databases used in this paper
and the web application providing on-line access to them were
constructed as part of the activities of the German Astrophysical
Virtual Observatory.
This work was partially supported by the Consejo de Investigaciones
Cient\'{\i}ficas y T\'ecnicas de la Rep\'ublica Argentina (CONICET)
and Secretar\'{\i}a de Ciencia y T\'ecnica, UNC (SeCyT).  CMdO
acknowledges financial help from FAPESP through the thematic project
01/07342-7.  
L.R.A. would like to thank FAPESP, CNPq and the Pr\'o-Reitoria de
Pesquisa, Universidade de S\~ao Paulo, for support.
\end{acknowledgements}

%%%%%%%%%%%%%%%%%%%%%%%%%%%%%%%%%%%%%%%%%%%%%%%%%%%%%%%%%%%%%%%%%%%%%%%%%%%
\bibliographystyle{aa} % style aa.bst
\bibliography{15347} % your references Yourfile.bib

\end{document}